\documentclass[12pt,preprintnumbers,aps,amssymb,nofootinbib,amsmath]
{revtex4}
\usepackage{epsfig,epsf}
\usepackage{graphicx}
\usepackage{epstopdf}
\newcommand{\beq}{\begin{equation}}
\newcommand{\eeq}{\end{equation}}
\def\eq#1{{(\ref{#1})}}
\def\fig#1{{Fig.~\ref{#1}}}

\newcommand{\be}{\begin{eqnarray}}
\newcommand{\ee}{\end{eqnarray}}

\newcommand{\as}{\alpha_s}
\newcommand{\un}{\underline}
\newcommand{\ben}{\begin{eqnarray*}}
\newcommand{\een}{\end{eqnarray*}}

\begin{document}

\preprint{{\bf Draft 1}}

~\vspace{1cm}

\title{\Large Production of $q$$\bar q$ Pairs in Proton--Nucleus 
Collisions\\ at High Energies\\[1cm]}

\author{Yuri V. Kovchegov$\,^a$\footnote{e-mail:
yuri@mps.ohio-state.edu} and Kirill Tuchin$\,^{b,c}$\footnote{e-mail:
tuchin@iastate.edu}}
\affiliation{$^a\,$Department of Physics, The Ohio State University,
Columbus, OH 43210 \\
$^b\,$Department of Physics and Astronomy, Iowa State University, Ames, IA 50011 \\
$^c\,$RIKEN BNL Research Center,
Upton, NY 11973-5000}

\date{\today}

\pacs{}

\begin{abstract} 
~\vspace*{5mm}\\ We calculate production of quark-antiquark pairs in
high energy proton-nucleus collisions both in the quasi-classical
approximation of McLerran-Venugopalan model and including quantum
small-$x$ evolution. The resulting production cross section is
explicitly expressed in terms of Glauber-Mueller multiple
rescatterings in the classical case and in terms of dipole-nucleus
scattering amplitude in the quantum evolution case. We generalize the
result of \cite{Tuchin:2004rb} beyond the aligned jet
configurations. We expand on the earlier results of Blaizot, Gelis and
Venugopalan \cite{Blaizot:2004wv} by deriving quark production cross
section including quantum evolution corrections in rapidity intervals
both between the quarks and the target and between the quarks and the
projectile.
\end{abstract}

\maketitle 

\thispagestyle{empty}

\newpage

\setcounter{page}{1}

%%%%%%%%%%%%%%%%%%%%%%%%%%%%%%%%%%%%%%%%%%%%%%%%%%%%%%%%%%%

\section{Introduction}

Heavy quark production in hadronic collisions in high energy QCD is
one of the most interesting and difficult problems. It is
characterized by two hard scales: heavy quark mass $m$ and the
saturation scale $Q_s$.  The threshold for the invariant mass of the
quark $q$ and antiquark $\bar q$ production is $2m$. Therefore, if $m$
is much larger than the confinement scale $\Lambda_{QCD}$, it
guarantees that a non-perturbative long distance physics has little
impact on the quark production \cite{Appelquist:tg} making
perturbative calculations possible \cite{collcharm} (for a review see
\cite{Brambilla:2004wf}).

Unlike the quark mass, which is a property of the produced quantum
state, the saturation scale $Q_s^2$ characterizes the density of color
charges in the wave function of each of the colliding hadrons
\cite{GLR,Mueller:wy,Blaizot:nc,MV}. It increases as a power of energy
and a power of atomic weight $A$ \cite{IV,Heri}.  At high energies and
especially in reactions with heavy nuclei it becomes significantly
larger than the confinement scale. It is the saturation scale which
makes the strong coupling constant small, $\as (Q_s) \ll 1$, insuring
applicability of the perturbative approach to all high energy
scattering problems \cite{MV}. For all processes involving heavy
quarks with momentum transfer of the order of $Q_s^2\sim m^2$ large
saturation scale implies breakdown of the collinear factorization
approach. The factorization approach may be extended by allowing the
incoming partons to be off-mass-shell.  This results in conjectured
$k_T$-factorization \cite{LRSS,CCH,CE}.  Although the phenomenological
applications of the $k_T$-factorization approach seem to be
numerically reasonable already at not very high energies
\cite{Fujii:2005vj} its theoretical status is not completely
justified. Like collinear factorization it is based on the leading
twist approximation. However, at sufficiently high energies, higher
twist contributions proportional to $(Q_s/m)^{2n}$ become important in
the kinematic region of small quark's transverse momentum, indicating
a breakdown of factorization approaches.

The fact that the saturation scale at high enough energies and for
large nuclei is large, $Q_s \gg \Lambda_{QCD}$, combined with the
observation that the typical transverse momentum of particles produced
in $pA$ scattering is of the order of that saturation scale, leads to
the conclusion that $Q_s$ sets the scale for the coupling constant,
making it small. This allows one to perform calculations for, say,
gluon production cross section in $pA$ collisions using the small
coupling approach \cite{KT}. The same line of reasoning can be applied
to heavy quark production considered here: the saturation scale $Q_s$
is the important hard scale making the coupling weak even if the quark
mass $m$ was small. Having the quark mass $m$ as another large
momentum scale in the problem only strengthens the case for
applicability of perturbative approach.

Resummation of leading higher twist corrections to all orders have
been performed before in the Color Glass Condensate (CGC)/saturation
framework for other observables not involving heavy quarks. The
problem of gluon production in $pA$ collisions at high energies was
solved by resuming the contributions which are enhanced by factors of
$\as^2 A^{1/3}\sim 1$ and $\as y\sim 1$, where $A$ is the atomic
number of the nucleus, and $y$ is the rapidity variable
\cite{KT}. Surprisingly, after resuming all such contributions to the 
single inclusive gluon production one recovers the $k_T$-factorization
formula \cite{KT} first suggested for the high parton density systems
in \cite{GLR}. Indeed, for large transverse momenta of the produced
gluons, $k_T \gg Q_s$, after neglecting all higher twist $(Q_s/k_T)^n$
corrections, one recovers the usual leading twist
$k_T$-factorization. It was quite amazing that $k_T$-factorization for
gluon production survived after resumming {\sl all twists} \cite{KT}.
However, $k_T$-factorization fails for the double inclusive gluon
production cross section \cite{JMK1}, as well as for the inclusive
quark production \cite{Fujii:2005vj}. Instead a more complicated
factorization picture emerges.

Indeed the fact that the produced gluon transverse momentum spectrum
in $pA$ collisions obtained in \cite{KT} still diverges proportional
to $\sim 1/k_T^2$ in the infrared introduces logarithmic dependence of
total gluon multiplicity $dN/dy$ (integrated over all transverse
momenta) on the infrared cutoff, raising questions about the
applicability of the perturbative approach for calculation of that
observable. However, while it is likely that $Q_s$ sets the scale for
the running coupling even in $dN/dy$, a formal analysis of the scale
of the running coupling is beyond the scope of this paper and is left
for future research. Similarly, if one is interested in obtaining
total quark multiplicity $dN_q/dy$ from the results presented below,
one should strictly speaking view them as derived for quark production
in deep inelastic scattering (DIS) (where the photon's virtuality $Q$
plays a role of the infrared cutoff keeping the physics perturbative),
which may also be applicable to $pA$ collisions.

Our goal in this paper is to calculate production of quark-antiquark
pairs in high energy proton-nucleus collisions and in DIS both in the
quasi-classical approximation of McLerran-Venugopalan model \cite{MV}
(summing powers of $\as^2 A^{1/3}$) and including quantum small-$x$
evolution (summing powers of $\as y $).  We generalize the result of
\cite{Tuchin:2004rb} for the single inclusive quark production beyond
the aligned jet configuration. We derive the double inclusive quark
and antiquark production.  We expand on the earlier results of
Blaizot, Gelis and Venugopalan \cite{Gelis:2003vh,Blaizot:2004wv} by
deriving a cross section that includes quantum corrections in the
rapidity intervals between the quarks and the target (powers of $\as
y$) and between the quarks and the projectile (powers of $\as
(Y-y)$). (Here $Y$ is the total rapidity interval, and $y$ is the
rapidity of the produced $q\bar q$ pair, with $0$ being the rapidity
of the target.) We generalize the approach of \cite{KopTar} by taking
into account valence quark rescatterings in the nucleus in the
quasi-classical approximation, and also by including the quantum
evolution corrections. In the quasi-classical limit our result should
be equivalent to solution of the Dirac equation in the background of
classical fields, similar to the one performed numerically in
\cite{Gelis:2004jp} for a collision of two nuclei.

The paper is structured as follows. We will first derive the $q\bar q$
production cross section in the quasi-classical approximation in
Sect. II. We will proceed by including quantum evolution corrections
in the obtained cross section in Sect. III. We will conclude in
sect. IV by discussing phenomenological applications of the obtained
results.

%%%%%%%%%%%%%%
\section{Inclusive Cross Section in the Quasi-Classical Approximation}

\begin{figure}[ht]
    \begin{center}
        \includegraphics[width=16cm]{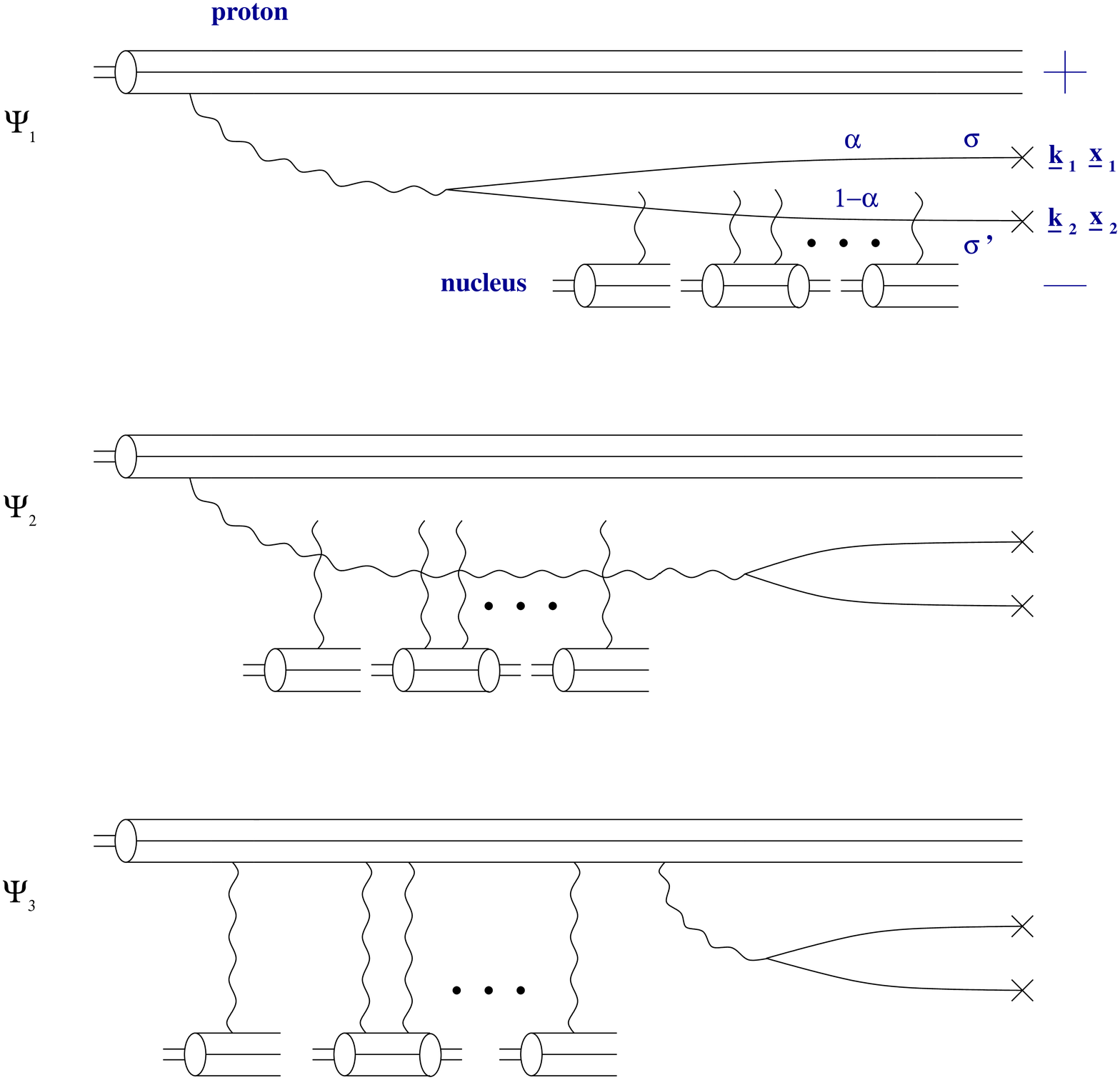}
\end{center}
\caption{Three main contributions to quark-antiquark production in 
the quasi-classical approximation.}
\label{wave_funct}
\end{figure}

The diagrams contributing to quark-antiquark pair production in the
quasi-classical approximation are shown in \fig{wave_funct}. The
graphs shown in \fig{wave_funct} are dominant in the light-cone gauge
of the proton. The first diagram corresponds to incoming valence quark
emitting a gluon, which splits into a quark-antiquark pair before the
system hits the target. The second diagram corresponds to the case
when the valence quark first emits a gluon, after which the system
rescatters on the target nucleus, and later the gluon splits into a
quark-antiquark pair. The third diagram corresponds to valence quark
rescattering on the target nucleus, after which it produces a gluon
which splits into a quark-antiquark pair.

The calculation of the diagrams in \fig{wave_funct} will proceed along
the lines outlined in \cite{KoM,KM} (see \cite{JMK} for a review)
using light cone perturbation theory \cite{BL}. In coordinate space
the diagram contributions factorize into a convolution of
Glauber-Mueller multiple rescattering with the ``wave function''
parts, which include splittings $q_v \rightarrow q_v\, G$ and $G
\rightarrow q\, \bar q$.

We begin by calculating the ``wave-function'' parts. In each of the
diagrams in \fig{wave_funct} they correspond to the two-step splitting
$q_v \rightarrow q_v\, G \rightarrow q_v\, q\, \bar q$. However, the
fact that the splittings take place either in initial or final states
depending on the diagram modifies the energy denominators, making the
``wave-function'' parts different in all three graphs. We will denote
these ``wave-function'' parts $\Psi_1$, $\Psi_2$ and $\Psi_3$
correspondingly, as shown in \fig{wave_funct}. The calculation of
$\Psi_1$, $\Psi_2$ and $\Psi_3$ proceeds according to the rules of
light cone perturbation theory (LCPT) \cite{BL} in the light cone
gauge of the proton, which we choose as moving in the light-cone
``plus'' direction (see \fig{wave_funct}). Calculations are first
performed in momentum space, after which the ``wave-functions'' are
Fourier-transformed into coordinate space.

The important subtlety of calculating final-state splittings is that
the light cone denominator for such splittings should be calculated
subtracting the light cone energy (the ``minus'' momentum component)
of the {\sl outgoing final} state. Indeed the light cone energies of
incoming and outgoing states are equal to each other: therefore, in
calculating final state splittings one can still subtract the incoming
energy in the denominators. However, in doing so one has to keep track
of a change in the minus component of the target's momentum, which
could be a bit tedious. For details on calculations of final state
emissions in the LCPT formalism see \cite{KT,JMK1}.

Since eikonal multiple rescatterings do not change the transverse
coordinates of the incoming quarks and the gluon, we can calculate
$\Psi_1$, $\Psi_2$ and $\Psi_3$ in transverse coordinate space by
calculating the diagrams in \fig{wave_funct} without interactions. We
assume that the outgoing quark and anti-quark have momenta $k_1$ and
$k_2$ correspondingly. The ``plus'' components of the momenta,
$k_{1+}$ and $k_{2+}$ are conserved in the interactions with the
target. Therefore, the ``plus'' component of the gluon's momentum is
equal to $k_{1+} + k_{2+}$. Here, for simplicity, we assume that
$k_{1+}, k_{2+} \ll p_+$, where $p_+$ is the typical light cone
momentum of the valence quarks in the proton. This implies that
$k_{1+} + k_{2+} \ll p_+$, i.e., that the gluon is also much softer
than the proton. In this kinematics the ``wave-functions'' in momentum
space are
\begin{subequations}\label{psi}
\begin{eqnarray}\label{psi1}
\Psi^{(1)}_{\sigma, \, \sigma'} (k_1,k_2) &=& 2 \, g \, T_a \, \sum_\lambda \, 
\frac{\un\epsilon^{*\lambda} \cdot (\un k_1 + \un k_2)}{(\un k_1 + \un k_2 )^2} \, 
g \, T_b \, \frac{1}{\frac{\un k_1^2 + m^2}{k_{1+}} + \frac{\un k_2^2
+ m^2}{k_{2+}}} \, \frac{\bar u_\sigma
(k_1)}{\sqrt{k_{1+}}} \gamma \cdot \epsilon^\lambda
\frac{v_{\sigma'}(k_2)}{\sqrt{k_{2+}}} \nonumber\\ 
&-& 2 \, g^2 \, T_a \, T_b \, \frac{1}{k_{1+} + k_{2+}} \,  \frac{1}{\frac{\un k_1^2 + m^2}{k_{1+}} + \frac{\un k_2^2
+ m^2}{k_{2+}}} \, \frac{\bar u_\sigma
(k_1)}{\sqrt{k_{1+}}} \gamma_+
\frac{v_{\sigma'}(k_2)}{\sqrt{k_{2+}}} \nonumber\\
&=& 2 \, g \, T_a
\, \sum_\lambda \, 
\frac{\un\epsilon^{*\lambda} \cdot (\un k_1 + \un k_2)}{(\un k_1 + \un
k_2 )^2} \, g \, T_b \, \frac{ L^\lambda_{\sigma, \, \sigma'}(\un k_1
\, (1-\alpha) - \un k_2 \, \alpha; \alpha)} 
{{\un k}_1^2 \, (1-\alpha) + {\un k}_2^2 \, \alpha + m^2}\,   \nonumber\\ 
&-& 4 \, g^2 \, T_a \, T_b \, \frac{\delta_{\sigma, \, \sigma'} \, \alpha \, (1-\alpha)}{{\un k}_1^2 \, (1-\alpha) + {\un k}_2^2 \, \alpha + m^2} ,
\end{eqnarray}
\begin{eqnarray}\label{psi2}
\Psi^{(2)}_{\sigma,\, \sigma'} (k_1,k_2) &=& 2 \, g \, T_a \, \sum_\lambda \, 
\frac{\un\epsilon^{*\lambda} \cdot (\un k_1 + \un k_2)}{(\un k_1 + \un k_2 )^2} \, 
g \, T_b \, \frac{1}{\frac{(\un k_1 + \un k_2
)^2} {k_{1+} + k_{2+}} -
\frac{\un k_1^2 + m^2}{k_{1+}} - \frac{\un k_2^2
+ m^2}{k_{2+}}} \,  \frac{\bar u_\sigma
(k_1)}{\sqrt{k_{1+}}} \gamma \cdot \epsilon^\lambda 
\frac{v_{\sigma'}(k_2)}{\sqrt{k_{2+}}} \nonumber\\   &=& - 2 \, g \, T_a
\, \sum_\lambda \, 
\frac{\un\epsilon^{*\lambda} \cdot (\un k_1 + \un k_2)}{(\un k_1 + \un
k_2 )^2} \, g \, T_b \, \frac{ L^\lambda_{\sigma, \, \sigma'}(\un k_1
\, (1-\alpha) - \un k_2 \, \alpha; \alpha)} 
{[\un k_1 \, (1-\alpha) - \un k_2 \, \alpha]^2 + m^2}\,,
\end{eqnarray}
\begin{eqnarray}\label{psi3}
\Psi^{(3)}_{\sigma, \, \sigma'} (k_1,k_2) &=& 2 \, g \, T_a  \sum_\lambda  
\frac{\un\epsilon^{*\lambda} \cdot (\un k_1 + \un k_2)}{- 
\frac{\un k_1^2 + m^2}{k_{1+}} - \frac{\un k_2^2
+ m^2}{k_{2+}}} \, g \, T_b \, \frac{(k_{1+} + k_{2+})^{-1}}{\frac{(\un k_1 + \un k_2
)^2}{k_{1+} + k_{2+}} -
\frac{\un k_1^2 + m^2}{k_{1+}} - \frac{\un k_2^2
+ m^2}{k_{2+}}} 
\frac{\bar u_\sigma (k_1)}{\sqrt{k_{1+}}} \, \gamma \cdot
\epsilon^\lambda \frac{v_{\sigma'}(k_2)}{\sqrt{k_{2+}}} \nonumber\\ 
&+& 2 \, g^2 \, T_a \, T_b \, \frac{1}{k_{1+} + k_{2+}} \,  \frac{1}{\frac{\un k_1^2 + m^2}{k_{1+}} + \frac{\un k_2^2
+ m^2}{k_{2+}}} \, \frac{\bar u_\sigma
(k_1)}{\sqrt{k_{1+}}} \gamma_+
\frac{v_{\sigma'}(k_2)}{\sqrt{k_{2+}}} \nonumber\\
&=& 2 \, g \,
T_a \sum_\lambda 
\frac{\un\epsilon^{*\lambda} \cdot (\un k_1 + \un k_2) \, \alpha \, (1-\alpha)}
{{\un k}_1^2 \, (1-\alpha) + {\un k}_2^2 \, \alpha + m^2} \, g \, T_b
\, \frac{ L^\lambda_{\sigma, \, \sigma'}(\un k_1
\, (1-\alpha) - \un k_2 \, \alpha; \alpha)} 
{[\un k_1 \, (1-\alpha) - \un k_2 \, \alpha]^2 + m^2}
\nonumber\\ &+& 4 \, g^2 \, T_a \, T_b \, \frac{\delta_{\sigma, \, \sigma'} \, \alpha \, (1-\alpha)}{{\un k}_1^2 \, (1-\alpha) + {\un k}_2^2 \, \alpha + m^2},
\end{eqnarray}
\end{subequations}
where \cite{KM}
\beq
L^\lambda_{\sigma, \, \sigma'}(\un k_1 \, (1-\alpha) - \un k_2 \, \alpha;
\alpha)\,=\, - \un \epsilon^\lambda \cdot [\un k_1 \, (1-\alpha) - \un k_2 \,
\alpha] \, (1-2 \, \alpha+\lambda \, \sigma) \, \delta_{\sigma,\, \sigma'}\,-\,
\frac{1}{\sqrt{2}} \, \sigma \, m \, (1-\lambda \, \sigma) \, 
\delta_{\sigma,-\sigma'}, 
\eeq
$\lambda = \pm 1$ is the gluon's polarization (which also does not
change under eikonal rescatterings), $\sigma = \pm 1$ and $\sigma' =
\pm1$ are quark and anti-quark helicities correspondingly (see
\fig{wave_funct}, $\sigma'$ is defined with respect to $-\vec k_2$), 
$m$ is the mass of the quarks, and the colors of
the gluon immediately after emission ($a$) and just before splitting
into $q\bar q$ pair ($b$) are kept different since the color of the
gluon is likely to change in interaction (for $\Psi_2$), due to which
the color factors will be calculated separately. Gluon polarization
vector for transverse gluons is given by $\epsilon_\mu^\lambda = (0,0, \un
\epsilon^\lambda)$, with $\un \epsilon^\lambda = (1 + i \, \lambda) /
\sqrt{2}$. The fraction of gluon's ``plus'' momentum carried by the
quark is denoted by $\alpha \equiv k_{1+} / (k_{1+} + k_{2+})$. The gluon 
"propagators" in diagrams $\Psi_1$ and $\Psi_3$ of \fig{wave_funct} have 
instantaneous (longitudinal) parts \cite{BL}, which account for the second (additive) 
terms in Eqs. (\ref{psi1}) and (\ref{psi3}).

Note that, as can be checked explicitly using \eq{psi}, 
\beq\label{neat}
\Psi^{(1)}_{\sigma, \, \sigma'} (k_1,k_2) + \Psi^{(2)}_{\sigma, \, 
\sigma'} (k_1,k_2) + \Psi^{(3)}_{\sigma, \, \sigma'} (k_1,k_2) 
\, = \, 0,
\eeq
indicating, of course, that there can be no emission without
interaction.

One may worry that since the gluon in the second graph of
\fig{wave_funct} interacts with the target, and, therefore the
interaction will depend on the transverse coordinate of this gluon,
instead of calculating $\Psi^{(2)}_{\sigma, \, \sigma'} (k_1,k_2)$ as
shown above in \eq{psi2}, one should separately calculate $q_v
\rightarrow q_v G$ and $G \rightarrow q \, \bar q$ transitions in 
momentum space, and then separately Fourier-transform each of the
results into coordinate space. However, this is not necessary, since
the gluon's transverse coordinate is uniquely fixed by the transverse
coordinates of the quark $\un x_1$ and the anti-quark $\un x_2$ and by
$\alpha$ (see e.g. \cite{KopTar,KNST,IKMT}). The gluon's transverse
coordinate is
\beq\label{u}
\un u \, = \, \alpha \, \un x_1 + (1-\alpha) \, \un x_2. 
\eeq
If we perform the calculations for $q_v
\rightarrow q_v G$ and $G \rightarrow q \, \bar q$ splittings independently, 
and Fourier-transform each of them into coordinate space, the $G
\rightarrow q \, \bar q$ component will come with a delta-function
$\delta^2 (\un u - \alpha \, \un x_1 + (1-\alpha) \, \un x_2)$, which
vanishes after integration over $\un u$ (which is an internal variable
and has to be integrated over) fixing $\un u$ at the value given by
\eq{u}. The result of this procedure is equivalent to a simple 
Fourier-transform of $\Psi^{(2)}_{\sigma, \, \sigma'} (k_1,k_2)$ from
\eq{psi2} into coordinate space.

The light cone ``wave-functions'' in transverse coordinate space are
defined as
\beq\label{mom.space}
\Psi^{(i)}_{\sigma,\, \sigma'} (\un x_1, \un x_2; \alpha) \,=\, 
\int \frac{d^2 k_1}{(2\pi)^2} \frac{d^2 k_2}{(2\pi)^2} \, e^{-i\un k_1 
\cdot \un x_1 - i \un k_2 \cdot \un x_2} \, \Psi^{(i)}_{\sigma,\, \sigma'} 
(k_1,k_2)\,,\quad i=1,2,3\,.
\eeq
Here we assume that the transverse coordinate of the valence quark
which emits the gluon (which splits into a $q\bar q$ pair) is $\un
0$. 

To perform the Fourier transform of \eq{mom.space} it is convenient to
introduce the following auxiliary functions
\begin{eqnarray}
F_2 (\un x_1, \un x_2; \alpha)&=& \int_0^\infty dq \, J_1 (q \, u) \,
K_1 \bigg(x_{12} \, \sqrt{m^2+
q^2\, \alpha(1-\alpha)} \bigg)\, \sqrt{m^2+
q^2\, \alpha(1-\alpha)} \,,\label{aux.fun1}\\ 
F_1 (\un x_1, \un x_2;\alpha)&=& \int_0^\infty dq \, J_1(q \, u) \, K_0 \bigg(x_{12}
\, \sqrt{m^2+q^2\, \alpha(1-\alpha)} \bigg)\,,\label{aux.fun2}\\
F_0 (\un x_1, \un x_2;\alpha)&=& \int_0^\infty dq \, q \, J_0(q \, u) \, K_0 \bigg(x_{12}
\, \sqrt{m^2+q^2\, \alpha(1-\alpha)} \bigg)\,,\label{aux.fun3}
\end{eqnarray}
where $u = |\un u|$, $\un x_{12} = \un x_1 - \un x_2$, $x_{12} =
|\un x_{12}|$, and $\un q = \un k_1 + \un k_2$. In terms of the functions 
defined in \eq{aux.fun1}, \eq{aux.fun2} and \eq{aux.fun3} we obtain
\begin{subequations}\label{msbar}
\begin{eqnarray}
\Psi^{(1)}_{\sigma, \, \sigma'} (\un x_1, \un x_2; \alpha) &=& 
\frac{2 \, g^2 \, T_a \, T_b}{(2 \pi)^2} \, \bigg\{ \, \sum_\lambda \,
 \bigg[ F_2 (\un x_1 ,\un x_2;\alpha) \, \frac{\un x_{12} \cdot 
\un\epsilon^\lambda}{x_{12}}\,(1- 2 \, \alpha +\lambda \, \sigma) 
\, \delta_{\sigma,\sigma'}
\nonumber\\
\, &+& \, F_1 (\un x_1 ,\un x_2; \alpha) \, \frac{i}{\sqrt{2}} \, 
\sigma \, m \, (1 - \lambda \, \sigma)
\, \delta_{\sigma, -\sigma'} \bigg] \, 
\frac{\un u\cdot \un \epsilon^{*\lambda}}{u}\, \nonumber\\
&-& 2 \, \delta_{\sigma, \, \sigma'} \, \alpha \, (1 - \alpha) \, 
F_0 (\un x_1 ,\un x_2; \alpha) \, \bigg\},\label{msbar1}\\
\Psi^{(2)}_{\sigma,\, \sigma'} (\un x_1, \un x_2; \alpha) &=& 
- \frac{2 \, g^2 \, T_a \, T_b}{(2\pi)^2} \, \sum_\lambda \,
\bigg[ \frac{\un x_{12} \cdot \un \epsilon^\lambda}{x_{12}} \, m \, 
K_1 (m\, x_{12}) \,(1-2\, \alpha + \lambda \, \sigma) \, 
\delta_{\sigma,\sigma'}
\nonumber\\
&& + \, K_0(m \, x_{12}) \, \frac{i}{\sqrt{2}} \, \sigma \, m \, (1
- \lambda \, \sigma) \,
\delta_{\sigma,-\sigma'}\bigg] \,\frac{\un
u \cdot \un \epsilon^{*\lambda}}{u^2}\,,\label{msbar2}\\
\Psi^{(3)}_{\sigma, \, \sigma'} (\un x_1, \un x_2; \alpha) &=&
- \Psi^{(1)}_{\sigma, \, \sigma'} (\un x_1, \un x_2; \alpha) 
- \Psi^{(2)}_{\sigma, \, \sigma'} (\un x_1, \un x_2; \alpha)\,.\label{msbar3}
\end{eqnarray}
\end{subequations}
The last relation \eq{msbar3} follows from \eq{neat}.

Summation over $\lambda$ yields
\begin{subequations}\label{ms}
\begin{eqnarray}
\Psi^{(1)}_{\sigma, \, \sigma'} (\un x_1, \un x_2; \alpha) &=& 
\frac{2 \, g^2 \, T_a \, T_b}{(2 \pi)^2} \,
\bigg[ F_2 (\un x_1 ,\un x_2;\alpha) \, \frac{1}{x_{12} \, u} \, 
[ (1- 2 \, \alpha) \, \un x_{12}\cdot \un u + i \, \sigma \, \epsilon_{ij} \,
u_i \, x_{12 \, j}] \, \delta_{\sigma,\sigma'}
\nonumber\\
&&\,+ \, F_1 (\un x_1 ,\un x_2; \alpha) \, \frac{i}{u} \, 
\sigma \, m \, (u_x + i \, \sigma \, u_y)
\, \delta_{\sigma, -\sigma'}  \,
\nonumber\\
&-& 2 \, \delta_{\sigma, \, \sigma'} \, \alpha \, (1 - \alpha) \, F_0 (\un x_1 ,\un x_2; \alpha) \, \bigg] ,\label{ms1}\\
\Psi^{(2)}_{\sigma,\, \sigma'} (\un x_1, \un x_2; \alpha) &=& 
- \frac{2 \, g^2 \, T_a \, T_b}{(2\pi)^2} \,
\bigg[ m \, K_1 (m\, x_{12}) \, \frac{1}{x_{12} \, u^2} \, 
[ (1- 2 \, \alpha) \, \un x_{12}\cdot \un u + i \, \sigma \, \epsilon_{ij} \,
u_i \, x_{12 \, j}] \, \delta_{\sigma,\sigma'}
\nonumber\\
&& + \, K_0(m \, x_{12}) \, \frac{i}{u^2} \, 
\sigma \, m \, (u_x + i \, \sigma \, u_y) \,
\delta_{\sigma,-\sigma'}\bigg] \,,\label{ms2}\\
\Psi^{(3)}_{\sigma, \, \sigma'} (\un x_1, \un x_2; \alpha) &=&
- \Psi^{(1)}_{\sigma, \, \sigma'} (\un x_1, \un x_2; \alpha) 
- \Psi^{(2)}_{\sigma, \, \sigma'} (\un x_1, \un x_2; \alpha)\,,\label{ms3}
\end{eqnarray}
\end{subequations}
where $\epsilon_{12} = 1 = - \epsilon_{21}$, $\epsilon_{11} =
\epsilon_{22} = 0$, and, assuming summation over repeating indices, 
$\epsilon_{ij} \, u_i \, v_j = u_x \, v_y - u_y \, v_x$. Also $x_{12
\, j}$ denotes the $j$th component of the vector $\un x_{12}$.

\begin{figure}
    \begin{center} \includegraphics[width=16.4cm]{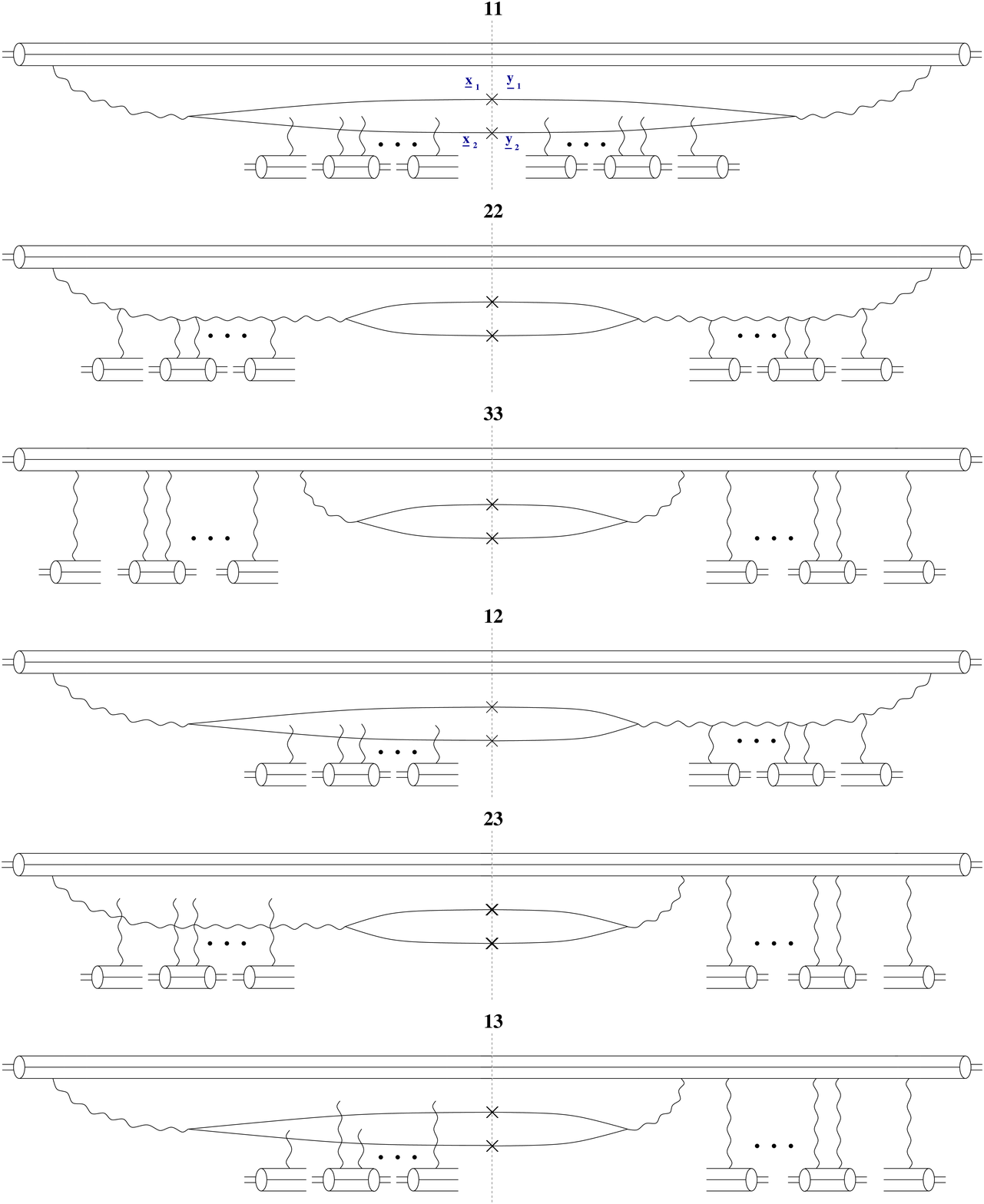}
\end{center}
\caption{Diagrams contributing to quark--anti-quark pair production in the 
quasi-classical approximation. Disconnected $t$-channel gluon lines
imply summation over all possible connections to the adjacent
$s$-channel quark and gluon lines.}
\label{xsect}
\end{figure}

Now that we have calculated the ``wave-functions'' in \eq{ms}, we can
proceed by calculating the $q\bar q$ production cross-section. The
relevant diagrams are shown in \fig{xsect} and are obtained by
squaring the sum of contributions from \fig{wave_funct}. We will first
calculate the parts of the diagrams in \fig{xsect} which are due to
the squares of the ``wave-functions'' from \eq{ms}. The resulting
expressions will then be convoluted with the multiple rescattering
parts of the diagrams.

The $q\bar q$ radiation kernel is obtained by averaging the square of
the sum of the ``wave functions'' from \eq{ms} over the quantum
numbers of the initial valence quark and summing over the quantum
numbers of the final state quarks. Since we are interested, first of
all, in the $q\bar q$ inclusive production cross section, where the
transverse momenta of both the quark $\un k_1$ and anti-quark $\un
k_2$ are fixed, in anticipation of a Fourier transform to transverse
momentum space, we will keep the transverse coordinates of the quarks
different in the amplitude and in the complex conjugate
amplitude. Therefore, if the transverse coordinates of the quarks are
$\un x_1$ and $\un x_2$ in the amplitude, we will denote them by $\un
y_1$ and $\un y_2$ in the complex conjugate amplitude, as shown in the
first graph of \fig{xsect}. The result for the squares of the
``wave-functions'' is
\beq\label{kernel}
\Phi_{ij}(\un x_1, \un x_2; \un y_1, \un y_2; \alpha) \, = \, 
\frac{1}{N_c}\sum_{\sigma,\sigma', a,b} \, 
\Psi^{(i)}_{\sigma, \, \sigma'} (\un x_1, \un x_2; \alpha)
\, \Psi^{(j) *}_{\sigma, \, \sigma'} (\un y_1, \un y_2; \alpha) \,,\quad i,j=1,2,3\,.
\eeq
Here the sum over gluons' colors $a$ and $b$ simply implies a
calculation of the color factors of the relevant diagrams, including
traces over fermion loops. Indeed these color factors, while
calculable in principle, are rather sophisticated, especially if we
are interested in the double-inclusive $q\bar q$ production cross
section. Therefore, to simplify the already quite involved
calculations, we will calculate the diagrams in \fig{xsect} in the
large-$N_c$ limit. The other reason for doing this is that, even
though it is clear how to improve on the large-$N_c$ approximation in
the classical limit, inclusion of quantum evolution beyond the
large-$N_c$ approximation would involve the functional JIMWLK \cite{JIMWLK}
evolution equation, a numerical solution of which is rather
involved. Therefore, in the calculations of ``wave-functions'' squared
below, the color factors will be calculated in the large-$N_c$ limit.

After a straightforward calculation we derive (we introduce the
gluon's transverse coordinate in the complex conjugate amplitude $\un
v \, \equiv \, \alpha \, \un y_1 + (1 - \alpha) \, \un y_2$ with $v =
|\un v|$ and $\un y_{12} = \un y_1 - \un y_2$, $y_{12} = |\un
y_{12}|$)
\begin{subequations}\label{phi}
\ben
\Phi_{11} (\un x_1, \un x_2; \un y_1,\un y_2 ; \alpha) \, = \, 4 \, C_F \, 
\bigg(\frac{\as}{\pi}\bigg)^2 \, \bigg\{ F_2(\un x_1, \un x_2; \alpha) \, 
F_2 (\un y_1, \un y_2; \alpha) \, \frac{1}{x_{12} \, y_{12} \, u \, v} \, 
[ (1 - 2 \, \alpha)^2 
\een
\ben
\times\,  (\un x_{12} \cdot \un u) \, (\un y_{12} \cdot \un v) 
+ (\epsilon_{ij} \, u_i \, x_{12 \, j}) \, (\epsilon_{kl} \, v_k \, y_{12 \, l})] +
F_1(\un x_1, \un x_2; \alpha) \, F_1(\un y_1, \un y_2; \alpha) \, m^2 \, 
\frac{\un u\cdot\un v}{u \, v} 
\een
\ben
+ 4 \, \alpha^2 \, (1-\alpha)^2 \,  F_0(\un x_1, \un x_2; \alpha) \, 
F_0(\un y_1, \un y_2; \alpha)  - 2 \, \alpha \, (1 - \alpha) \, (1 - 2 \, \alpha) \,
\bigg[ \frac{\un x_{12} \cdot \un u}{x_{12} \, u} \, F_2(\un x_1, \un x_2; \alpha) 
\een
\be\label{phi11}
\times \,
F_0(\un y_1, \un y_2; \alpha) +  
\frac{\un y_{12} \cdot \un v}{y_{12} \, v} \, 
F_2(\un y_1, \un y_2; \alpha) \, F_0(\un x_1, \un x_2; \alpha) \bigg] \, \bigg\} \,,
\ee
%%%%%%%%%%%%%%%%%%%%%%%%%%%%%%%%%%%%%%%%%%%%%%%%%%%%%%%
\ben
\Phi_{22} (\un x_1, \un x_2; \un y_1,\un y_2; \alpha) \, = \, 4 \, C_F \,
\bigg(\frac{\as}{\pi}\bigg)^2 \, m^2 \, \bigg\{
K_1(m \, x_{12}) \, K_1(m \, y_{12}) \, \frac{1}{x_{12} \, y_{12} \, u^2 \, v^2} 
\, [ (1 - 2 \, \alpha)^2
\een
\be\label{phi22}
\times\,(\un x_{12} \cdot \un u) \, (\un y_{12} \cdot \un v) 
+ (\epsilon_{ij} \, u_i \, x_{12 \, j}) \, (\epsilon_{kl} \, v_k \, y_{12 \, l})] +
K_0(m \, x_{12})\, K_0(m \, y_{12}) \, \frac{\un u\cdot\un v}{u^2 \, v^{2}} \bigg\}\,,
\ee
%%%%%%%%%%%%%%%%%%%%%%%%%%%%%%%%%%%%%%%%%%%%%%%%%%%%%%%%%%%%%%
\ben
\Phi_{12} (\un x_1, \un x_2; \un y_1, \un y_2; \alpha) = -  4 \, C_F \, 
\bigg(\frac{\as}{\pi}\bigg)^2 \, m \,
\bigg\{ F_2 (\un x_1, \un x_2;\alpha) \, K_1(m \, y_{12}) \, 
\frac{1}{x_{12} \, y_{12} \, u \, v^2} \, [ (1 - 2 \, \alpha)^2
\een
\ben
\times\, (\un x_{12} \cdot \un u) \, (\un y_{12} \cdot \un v) 
+ (\epsilon_{ij} \, u_i \, x_{12 \, j}) \, (\epsilon_{kl} \, v_k \, y_{12 \, l})]
+ m \, F_1(\un x_1,\un x_2; \alpha) 
\, K_0(m \, y_{12}) \, \frac{\un u\cdot\un v}{u \, v^{2}} 
\een
\be\label{phi12}
- 2 \,  \alpha \, (1 - \alpha) \, (1 - 2 \, \alpha) \,
\frac{\un y_{12} \cdot \un v}{y_{12} \, v^2}  \, 
F_0(\un x_1, \un x_2; \alpha) \, K_1 (m \, y_{12}) \bigg\}\,, 
\end{eqnarray}
\ben
\Phi_{33} (\un x_1, \un x_2; \un y_1, \un y_2; \alpha) \, =\, 
\Phi_{11} (\un x_1, \un x_2; \un y_1, \un y_2; \alpha) + 
\Phi_{22} (\un x_1, \un x_2; \un y_1, \un y_2; \alpha) +
\Phi_{12} (\un x_1, \un x_2; \un y_1, \un y_2; \alpha) 
\een
\be\label{phi33}
+ \Phi_{21} (\un x_1, \un x_2; \un y_1, \un y_2; \alpha)
\ee
\be\label{phi13}
\Phi_{13} (\un x_1, \un x_2; \un y_1, \un y_2; \alpha) \, =\, 
- \Phi_{11} (\un x_1, \un x_2; \un y_1, \un y_2; \alpha) 
- \Phi_{12} (\un x_1, \un x_2; \un y_1, \un y_2; \alpha)
\ee
\be\label{phi23}
\Phi_{23} (\un x_1, \un x_2; \un y_1, \un y_2; \alpha) \, =\, 
- \Phi_{21} (\un x_1, \un x_2; \un y_1, \un y_2; \alpha) 
- \Phi_{22} (\un x_1, \un x_2; \un y_1, \un y_2; \alpha)
\ee
\be\label{phi_symm}
\Phi_{ij} (\un x_1, \un x_2; \un y_1, \un y_2; \alpha) \, =\, 
\Phi_{ji}^* (\un y_1, \un y_2; \un x_1, \un x_2; \alpha).
\ee
\end{subequations}
Here Eqs. (\ref{phi33},\ref{phi13},\ref{phi23}) follow from
\eq{ms3}. \eq{phi_symm} allows one to obtain $\Phi_{21}$,  
$\Phi_{31}$ and $\Phi_{32}$ from \eq{phi12}, \eq{phi13} and
\eq{phi23}.

Rescattering of $q_v$, $q_v \, G$ and $q_v\, q\, \bar q$
configurations on a large nucleus brings in different factors, which
we label $\Xi_{ij}$ depending on the diagram shown in
\fig{xsect}. For the case of single-quark inclusive production 
cross section (when transverse momentum of one of the quarks is
integrated over) they were calculated in \cite{Tuchin:2004rb}. As we
mentioned above, the calculations complicate tremendously for the
double-inclusive $q\bar q$ production cross section we are interested
in calculating here. We will, therefore, perform our calculations on
the large-$N_c$ limit. Introducing quark saturation scale \cite{KM,JMK}
\beq
Q_s^2 \, = \, \frac{4 \, \pi \, \as^2 \, C_F}{N_c} \, \rho \, T(\un b)
\, \approx \, 2 \, \pi \, \as^2 \, N_c \, \rho \, T(\un b)
\eeq
with $\rho$ the nucleon number density in the nucleus and $T(\un b)$
the nuclear profile function, we write
\begin{subequations}\label{xi}
\begin{eqnarray}
\Xi_{11} (\un x_1, \un x_2; \un y_1, \un y_2; \alpha) 
&=&e^{-\frac{1}{4}\, (\un x_1 -\un y_1)^2 \, Q_s^2 \, \ln 
(1/|\un x_1 -\un y_1| \, \Lambda)
-\frac{1}{4}\, (\un x_2 -\un y_2)^2 \, Q_s^2 \, 
\ln (1/|\un x_2 -\un y_2| \, \Lambda)}\,,\\
\Xi_{22} (\un x_1, \un x_2; \un y_1, \un y_2; \alpha) 
&=&e^{-\frac{1}{2}\, (\un u-\un v)^2 \, Q_s^2 \, 
\ln (1/ |\un u -\un v| \, \Lambda)}\,,\\
\Xi_{33} (\un x_1, \un x_2; \un y_1, \un y_2; \alpha) &=& 1\,,\\
\Xi_{12} (\un x_1, \un x_2; \un y_1, \un y_2; \alpha) &=& e^{-\frac{1}{4}\, 
(\un x_1 -\un v)^2 \, Q_s^2 \, 
\ln (1/ |\un x_1 -\un v| \, \Lambda) -\frac{1}{4}\, (\un x_2 -\un v)^2 \, Q_s^2 \, 
\ln (1/ |\un x_2 -\un v| \, \Lambda)}\, ,\\
\Xi_{23} (\un x_1, \un x_2; \un y_1, \un y_2; \alpha) &=& e^{-\frac{1}{2} \, u^2 \, 
Q_s^2 \, \ln (1/u \, \Lambda)}\,,\\
\Xi_{13} (\un x_1, \un x_2; \un y_1, \un y_2; \alpha) &=& e^{-\frac{1}{4} \, x_1^2 \, 
Q_s^2 \, \ln (1/x_1 \Lambda) - \frac{1}{4} \, x_2^2 \, Q_s^2 \, \ln (1/x_2 \Lambda)}\,
\end{eqnarray}
\end{subequations} 
with $\Lambda$ some infrared cutoff. All other $\Xi_{ij}$'s can be
found from the components listed in
\eq{xi} using
\beq
\Xi_{ij} (\un x_1, \un x_2; \un y_1, \un y_2; \alpha) \, = \, 
\Xi_{ji} (\un y_1, \un y_2; \un x_1, \un x_2; \alpha)
\eeq
similar to \eq{phi_symm}. Note that in arriving at equations \eq{xi} we have used 
the fact that the valence quark rescatterings on the target cancel due 
to real-virtual cancellations in the first four graphs in \fig{xsect} \cite{KoM}. 
Such cancellations do not happen completely for a projectile dipole, as we will see 
in Section III. 

Using Eqs. (\ref{phi}) and (\ref{xi}) we derive the double-inclusive
quark--anti-quark production cross section in $pA$ collisions in the
quasi-classical approximation
\ben
\frac{d \, \sigma}{d^2k_1 \, d^2 k_2 \, dy \, d \alpha \, d^2 b} \,=\, 
\frac{1}{4 \, (2 \, \pi)^6} \, \int d^2 x_1 \, d^2 x_2 \, d^2 y_1 \, d^2 y_2 \, 
e^{- i \, \un k_1 \cdot (\un x_1 - \un y_1) - i \, \un k_2 \cdot (\un x_2 - \un y_2)} 
\een
\beq\label{dcl}
\times \, \sum_{i,j=1}^3\, \Phi_{ij} (\un x_1, \un x_2; \un y_1, \un y_2; \alpha) \
\Xi_{ij} (\un x_1, \un x_2; \un y_1, \un y_2; \alpha).
\eeq
Here $y$ is the rapidity of the $s$-channel gluon, which splits into
the $q\bar q$ pair. Since the quark and the anti-quark are most likely
to be produced close to each other in rapidity, one can think of $y$
as the rapidity of the quarks. $\un b$ is the impact parameter of the
proton with respect to the nucleus.

The single inclusive quark production cross section is easily obtained
from \eq{dcl} by integrating over one of the quark's momenta
\ben
\frac{d \, \sigma}{d^2 k \, dy \, d^2 b} \,=\, \frac{1}{2 \, (2 \, \pi)^4} \, \int  
d^2 x_1 \, d^2 x_2 \, d^2 y_1 \, \int_0^1 d\alpha \, e^{-i \, \un k \cdot (\un x_1-\un
y_1)} \, \sum_{i,j=1}^3 \, \Phi_{ij}\, (\un x_1, \un x_2; \un y_1, \un
x_2; \alpha) 
\een
\beq\label{single_cl}
\times \, \Xi_{ij} (\un x_1, \un x_2; \un y_1, \un x_2; \alpha)\,,
\eeq
where we inserted an overall factor of $2$ to account for both quarks
and anti-quarks. In \eq{single_cl} $y$ is the rapidity of the produced
(anti-)quark.

%%%%%%%%%%%%%%%%%%%%%%%%%%%%%%%%%%%%%%%%%%%%%%%%%%%%%%%%%%%%%%%%%%%%%%%%%%%%

\section{Including Quantum Evolution}

Here we are going to include small-$x$ nonlinear quantum evolution of
\cite{BK} into the cross sections from Eqs. (\ref{dcl}) and (\ref{single_cl}).  
Since the evolution equations in \cite{BK} are written for the forward
amplitude of a quark dipole on a nucleus, we have to first generalize
\eq{dcl} to the case of $q\bar q$ production in dipole-nucleus scattering. 
Indeed, strictly speaking our results would then only be applicable to
particle production in deep inelastic scattering. However, our results
below may still serve as a good approximation for gluon production in
$pA$ collisions \cite{KT}.  The generalization of Eqs. (\ref{dcl}) and
(\ref{single_cl}) to dipole-nucleus scattering is easily done by
including emissions of the $s$-channel gluon in \fig{xsect} by the
quark and anti-quark in the incoming dipole. If the transverse
coordinates of the quark and anti-quark in the incoming dipole are
denoted by $\un z_0$ and $\un z_1$ correspondingly with $\un z_{01} =
\un z_0 - \un z_1$, we write
\ben
\frac{d \, \sigma}{d^2k_1 \, d^2 k_2 \, dy \, d \alpha \, d^2 b} (\un z_{01}) \,=\, 
\frac{1}{4 \, (2 \, \pi)^6} \, \int d^2 x_1 \, d^2 x_2 \, d^2 y_1 \, d^2 y_2 \, 
e^{- i \, \un k_1 \cdot (\un x_1 - \un y_1) - i \, \un k_2 \cdot (\un x_2 - \un y_2)} 
\een
\beq\label{dcl_dip}
\times \, \sum_{i,j=1}^3\, \sum_{k,l=0}^1 \, (-1)^{k+l} \, \Phi_{ij} (\un x_1 - \un z_k, 
\un x_2 - \un z_k; \un y_1 - \un z_l, \un y_2 - \un z_l; \alpha) \
\Xi_{ij} (\un x_1, \un x_2, \un z_k; \un y_1, \un y_2, \un z_l; \alpha),
\eeq
where now we have
\begin{subequations}\label{xi_dip}
\begin{eqnarray}
\Xi_{11} (\un x_1, \un x_2, \un z_k; \un y_1, \un y_2, \un z_l; \alpha) 
&=&e^{-\frac{1}{4}\, (\un x_1 -\un y_1)^2 \, Q_s^2 \, \ln 
(1/|\un x_1 -\un y_1| \, \Lambda)
-\frac{1}{4}\, (\un x_2 -\un y_2)^2 \, Q_s^2 \, \ln 
(1/|\un x_2 -\un y_2| \, \Lambda)}\,,\\
\Xi_{22} (\un x_1, \un x_2, \un z_k; \un y_1, \un y_2, \un z_l; \alpha) 
&=&e^{-\frac{1}{2}\, (\un u-\un v)^2 \, Q_s^2 \, 
\ln (1/ |\un u -\un v| \, \Lambda)}\,,\\
\Xi_{33} (\un x_1, \un x_2, \un z_k; \un y_1, \un y_2, \un z_l; \alpha) &=& 
e^{- \frac{1}{2} z_{kl}^2 Q_s^2 \, \ln (1/ z_{kl} \, \Lambda) }\,,\\
\Xi_{12} (\un x_1, \un x_2, \un z_k; \un y_1, \un y_2, \un z_l; \alpha) &=& 
e^{-\frac{1}{4}\, (\un x_1 -\un v)^2 \, Q_s^2 \, 
\ln (1/ |\un x_1 -\un v| \, \Lambda) -\frac{1}{4}\, (\un x_2
-\un v)^2 \, Q_s^2 \, 
\ln (1/ |\un x_2 -\un v| \, \Lambda)}\, ,\\
\Xi_{23} (\un x_1, \un x_2, \un z_k; \un y_1, \un y_2, \un z_l; \alpha) &=& 
e^{-\frac{1}{2} \, (\un u - \un z_l)^2 \, Q_s^2 \, 
\ln (1/ |\un u - \un z_l| \, \Lambda)}\,,\\
\Xi_{13} (\un x_1, \un x_2, \un z_k; \un y_1, \un y_2, \un z_l; \alpha) &=& 
e^{-\frac{1}{4} \, (\un x_1 - \un z_l)^2 \, Q_s^2 \, 
\ln (1/ |\un x_1 - \un z_l| \, \Lambda) - \frac{1}{4} \, 
(\un x_2 - \un z_l)^2 \, Q_s^2 \, 
\ln (1/ |\un x_2 - \un z_l| \, \Lambda)}\,.
\end{eqnarray}
\end{subequations} 
Again, all other $\Xi_{ij}$'s can be found from the components listed in
\eq{xi_dip} using
\beq
\Xi_{ij} (\un x_1, \un x_2, \un z_k; \un y_1, \un y_2, \un z_l; \alpha) \, = \, 
\Xi_{ji} (\un y_1, \un y_2, \un z_l; \un x_1, \un x_2, \un z_k; \alpha).
\eeq

The inclusion of quantum corrections due to leading logarithmic
(resumming powers of $\as \, y$) approximation in the large-$N_c$
limit is done along the lines of \cite{KT} (see also
\cite{JMK1,Marquet,yuri05} and \cite{JMK} for a review) using Mueller's dipole 
model formalism \cite{dipole}. Since the integration over rapidity
interval separating the quark and the anti-quark in the pair does not
generate a factor of the total rapidity interval $Y$ of the collision
(i.e., does not give a leading logarithm of energy), the prescription
for inclusion of quantum evolution is identical to the single gluon
production case. We first define the quantity $n_1 ({\un z}_{0}, \un
z_1; {\un w}_{0}, \un w_{1}; Y-y)$, which has the meaning of the
number of dipoles with transverse coordinates ${\un w}_{0}, \un
w_{1}$ at rapidity $y$ generated by the evolution from the original
dipole ${\un z}_{0}, \un z_1$ having rapidity $Y$. It obeys the dipole
equivalent of the BFKL evolution equation \cite{dipole,BFKL}
\ben
\frac{\partial n_1 ({\un z}_{0}, \un
z_1; {\un w}_{0}, \un w_{1}; y)}{\partial y} \, = \, 
\frac{\as \, N_c}{2 \, \pi^2} \, 
\int d^2 z_2 \, \frac{z_{01}^2}{z_{20}^2 \, z_{21}^2} \, 
\bigg[ n_1 ({\un z}_{0}, \un
z_2; {\un w}_{0}, \un w_{1}; y) + 
n_1 ( {\un z}_{2}, \un
z_1; {\un w}_{0}, \un w_{1}; y) 
\een
\be\label{eqn}
- n_1 ({\un z}_{0}, \un z_1; {\un w}_{0}, \un w_{1}; y) \bigg] 
\ee
with the initial condition 
\be
n_1 ({\un z}_{0}, \un z_1; {\un w}_{0}, \un w_{1}; y=0) \, = \,
\delta ( \un z_0 - \un w_{0} ) \, \delta ( \un z_1 - \un w_{1} ).
\ee
If the target nucleus has rapidity $0$, the incoming dipole has
rapidity $Y$, and the produced quarks have rapidity $y$, the inclusion
of small-$x$ evolution in the rapidity interval $Y-y$ is accomplished
by replacing the cross section from \eq{dcl_dip} by \cite{KT,Marquet,JMK}
\beq\label{sub1}
\frac{d \, \sigma}{d^2k_1 \, d^2 k_2 \, dy \, d \alpha \, d^2 b} (\un z_{01}) 
\rightarrow \int d^2 w_{0} \, d^2 w_{1} \, n_1 ({\un z}_{0}, \un z_1; 
{\un w}_{0}, \un w_{1}; Y-y) \, 
\frac{d \, \sigma}{d^2k_1 \, d^2 k_2 \, dy \, d \alpha \, d^2 b} (\un w_{01}).
\eeq
Note that while the substitution in \eq{sub1} includes only linear
evolution, it results from analyzing all the possible non-linear
evolution corrections including all possible pomeron splittings
between the projectile and the produced $q\bar q$ pair. As was
originally shown in \cite{KT} the pomeron splittings cancel in the
rapidity interval between $y$ and $Y$, leaving only the linear
evolution contribution included in \eq{sub1}.

Inclusion of evolution in the interval between $0$ and $y$ is
accomplished by replacing the Mueller-Glauber rescattering exponents
according to the following rule \cite{KT}
\beq\label{sub2}
e^{- \frac{1}{4} \, (\un x_0 - \un x_1)^2 \, Q_s^2 \, 
\ln (1/|\un x_0 - \un x_1| \, \Lambda)} \, \rightarrow \,
1 - N (\un x_0, \un x_1, Y)
\eeq
where $N (\un x_0, \un x_1, Y)$ is the forward amplitude for a quark
dipole $\un x_0, \un x_1$ scattering on a target with rapidity
interval $Y$ between the dipole and the target. It obeys the following
evolution equation \cite{BK}
\ben
\frac{\partial N ({\un x}_{0}, {\un x}_1, Y)}{\partial Y} \, = \, 
\frac{\as \, N_c}{2 \, \pi^2} \, 
\int d^2 x_2 \, \frac{x_{01}^2}{x_{20}^2 \, x_{21}^2} \, 
\left[ N ({\un x}_{0}, {\un x}_2, Y) + 
N ({\un x}_{2}, {\un x}_{1}, Y) - N ({\un
x}_{0}, {\un x}_1, Y) \right. 
\een
\beq\label{eqN}
- \left. N ({\un x}_{0}, {\un x}_{2}, Y) \, N ({\un x}_{2}, {\un x}_{1}, Y)
\right]
\eeq
with the initial condition
\beq
 N ({\un x}_{0}, {\un x}_1, Y=0) \, = \, 1 - e^{- \frac{1}{4} \, (\un
 x_0 - \un x_1)^2 \, Q_s^2 \, \ln (1/|\un x_0 - \un x_1| \, \Lambda)}.
\eeq

Performing the substitution from \eq{sub2} in \eq{xi_dip} yields
\begin{subequations}\label{xi_ev}
\be
\Xi_{11} (\un x_1, \un x_2, \un z_k; \un y_1, \un y_2, \un z_l; \alpha, Y) 
&=& - N (\un x_1 , \un y_1, Y)  - N (\un x_2 , \un y_2, Y) \nonumber \\ && 
+ N (\un x_1 , \un y_1, Y) \,  N (\un x_2 , \un y_2, Y) \, , \\
\Xi_{22} (\un x_1, \un x_2, \un z_k; \un y_1, \un y_2, \un z_l; \alpha, Y) 
&=& - 2 \, N (\un u, \un v, Y) + N (\un u, \un v, Y)^2 \,,\\
\Xi_{33} (\un x_1, \un x_2, \un z_k; \un y_1, \un y_2, \un z_l; \alpha, Y) &=& 
- 2 \, N (\un z_k, \un z_l, Y) + N (\un z_k, \un z_l, Y)^2 \,,\\
\Xi_{12} (\un x_1, \un x_2, \un z_k; \un y_1, \un y_2, \un z_l; \alpha, Y) &=& 
- N (\un x_1 , \un v, Y)  - N (\un x_2 , \un v, Y)  \nonumber \\ &&
+ N (\un x_1 , \un v, Y) \,  N (\un x_2 , \un v, Y) \, ,\\
\Xi_{23} (\un x_1, \un x_2, \un z_k; \un y_1, \un y_2, \un z_l; \alpha, Y) &=& 
- 2 \, N (\un u, \un z_l, Y) + N (\un u, \un z_l, Y)^2 \,,\\
\Xi_{13} (\un x_1, \un x_2, \un z_k; \un y_1, \un y_2, \un z_l; \alpha, Y) &=& 
- N (\un x_1 , \un z_l, Y)  - N (\un x_2 , \un z_l, Y)  \nonumber \\ &&
+ N (\un x_1 , \un z_l, Y) \,  N (\un x_2 , \un z_l, Y) \,,
\end{eqnarray}
\end{subequations}
with 
\beq
\Xi_{ij} (\un x_1, \un x_2, \un z_k; \un y_1, \un y_2, \un z_l; \alpha, Y) \, = \, 
\Xi_{ji} (\un y_1, \un y_2, \un z_l; \un x_1, \un x_2, \un z_k; \alpha, Y).
\eeq
In arriving at \eq{xi_ev} we have dropped additive unit terms which do
not contribute to the cross-section due to \eq{neat} leading to 
$\sum_{ij=1}^3 \Phi_{ij} =0$.

With the definition of Eqs. (\ref{xi_ev}) we write the following
answer for the double inclusive $q\bar q$ production cross section
including small-$x$ evolution effects
\ben
\frac{d \, \sigma}{d^2 k_1 \, d^2 k_2 \, dy \, d \alpha \, d^2 b} (\un z_{01}) 
\, = \, \frac{1}{4 \, (2 \, \pi)^6} \, 
\int d^2 w_{0} \, d^2 w_{1} \, n_1 ({\un z}_{0}, \un z_1; 
{\un w}_{0}, \un w_{1}; Y-y) \,
\een
\ben
\times \,  d^2 x_1 \, d^2 x_2 \, d^2 y_1 \, d^2 y_2 \, 
e^{- i \, \un k_1 \cdot (\un x_1 - \un y_1) - i \, \un k_2 \cdot (\un x_2 - \un y_2)} 
\een
\beq\label{dcl_ev}
\times \, \sum_{i,j=1}^3\, \sum_{k,l=0}^1 \, (-1)^{k+l} \, 
\Phi_{ij} (\un x_1 - \un w_k, 
\un x_2 - \un w_k; \un y_1 - \un w_l, \un y_2 - \un w_l; \alpha) \
\Xi_{ij} (\un x_1, \un x_2, \un w_k; \un y_1, \un y_2, \un w_l; \alpha, y).
\eeq
Similar to how we arrived at \eq{single_cl}, we integrate over one of
the quarks' transverse momenta to obtain the single inclusive quark
production cross section
\ben
\frac{d \, \sigma}{d^2 k \, \, dy \, d^2 b} (\un z_{01}) 
\, = \, \frac{1}{2 \, (2 \, \pi)^4} \, 
\int d^2 w_{0} \, d^2 w_{1} \, n_1 ({\un z}_{0}, \un z_1; 
{\un w}_{0}, \un w_{1}; Y-y) \, d^2 x_1 \, d^2 x_2 \, d^2 y_1 \,  
e^{- i \, \un k \cdot (\un x_1 - \un y_1) }
\een
\beq\label{single_ev}
\times \int_0^1 d\alpha \sum_{i,j=1}^3 \sum_{k,l=0}^1 (-1)^{k+l} \, 
\Phi_{ij} (\un x_1 - \un w_k, 
\un x_2 - \un w_k; \un y_1 - \un w_l, \un x_2 - \un w_l; \alpha) \
\Xi_{ij} (\un x_1, \un x_2, \un w_k; \un y_1, \un x_2, \un w_l; \alpha, y).
\eeq
Equations \eq{dcl_ev} and \eq{single_ev} are the central results of
this paper. 

\begin{figure}[htb]
    \begin{center}
        \includegraphics[width=10cm]{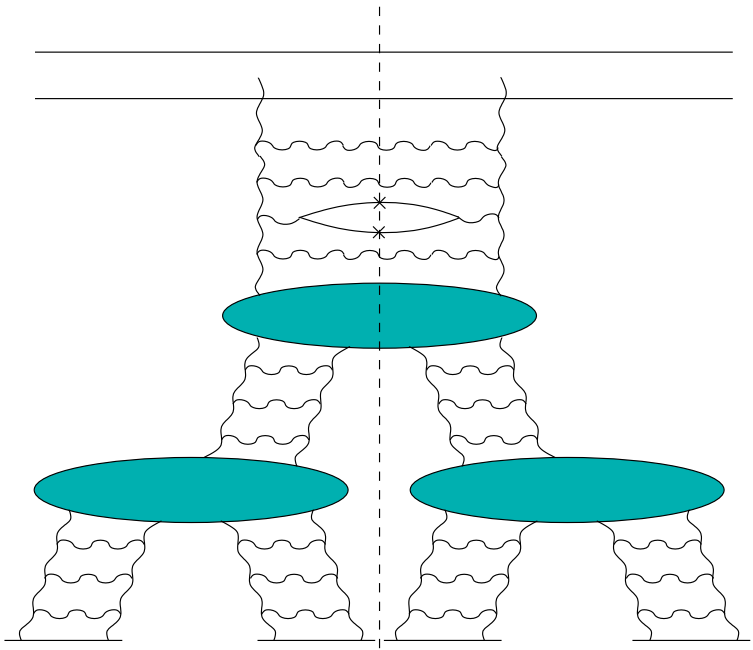}
\end{center}
\caption{An example of a pomeron fan diagram contributing to the $q\bar q$ 
  production cross section in proton-nucleus collisions or in DIS as
  calculated in equations \eq{dcl_ev} and \eq{single_ev}. The produced
  quark and anti-quark are marked by crosses.}
\label{fans}
\end{figure}

An example of the pomeron fan diagram contributing to the obtained
cross sections is shown in \fig{fans}. There the proton or $q\bar q$
dipole in DIS is shown on top of the figure. The nucleus is
represented by the straight lines at the bottom of the figure. As
usual each ladder represents a BFKL pomeron. \fig{fans} reflects the
main features of equations \eq{dcl_ev} and \eq{single_ev}: it contains
a linear evolution between the produced $q\bar q$ pair and the
projectile, and the non-linear evolution between the $q\bar q$ pair
and the target. Indeed a relatively simple diagram in \fig{fans} does
not include all the complexity of the nonlinear interactions in
\eq{xi_ev} and of the emission wave functions in \eq{phi}.
%%%%%%%%%%%%%%%%%%%%%%%%%%%%%%%%%%%%%%%%%%%%%%%%%%%%%%%%%%%%%%%%%%%%
%%%%%%%%%%%%%%%%%%%%%%%%%%%%%%%%%%%%%%%%%%%%%%%%%%%%%%%%%%%%%%%%%%%%

%%%%%%%%%%%%
\section{Summary}

Expressions \eq{dcl_ev} and \eq{single_ev} for the single and double
inclusive quark production have been derived by summing perturbation
series in the coupling constant $\as$. In that sense our result is
perturbative. It was pointed out in
\cite{Kharzeev:2005iz,Kharzeev:2006zm, Gelis:2005gs} that there can be
a significant non-perturbative contribution to particle production in
high energy QCD. The investigation of this effect is beyond the scope
of the present paper: however it certainly deserves further study.
 
Equations \eq{dcl_ev} and \eq{single_ev}  have important
phenomenological applications for studying the dense partonic system
in p(d)A and eA collisions. Observation of hadron suppression in the
nuclear modification factor measured in dA collisions at forward
rapidities at Relativistic Heavy Ion Collider (RHIC) \cite{dAdata}
signals the onset of the nonlinear evolution of the scattering
amplitude for light hadrons \cite{KLM,Kharzeev:2003wz,KW}. Due to a
large mass, the impact of nonlinear evolution effects on the heavy
quark production is shifted to higher energy and/or rapidity. It was
estimated in \cite{KhT} using the $k_T$-factorization approach that
one can expect a significant deviation of the open charm production
cross section from the perturbative behavior already at
pseudo-rapidity $\eta\simeq 2$ at RHIC. Due to the heavy quark
production threshold one expects that the total multiplicity of
open charm scales as $N_\mathrm{coll}$ at lower energy and/or
rapidity whereas at higher energies and/or rapidities the scaling
law should coincide with that for lighter hadrons \cite{KhT}, i.\
e.\ open charm multiplicity should scale as $N_\mathrm{part}$
\cite{KL} due to high parton density effects.  Therefore, to be able
to compare predictions of CGC with the data reported by RHIC
experiments and to make predictions for the possible upcoming $pA$ run
at the Large Hadron Collider (LHC), it is important to perform a
calculation of an open charm production within the more general
approach developed in this paper. Our final results
\eq{dcl_ev} and \eq{single_ev} allow one to describe open charm
transverse momentum spectra at different rapidities and center-of-mass
energies, allowing for a complete description of RHIC and LHC
data. Since the saturation scale $Q_s$ is expected to be even higher
at LHC than it was at RHIC, the CGC effects on heavy quark production
at LHC should be even more significant.

%%%%%%%%%%%%%%%%%%%%%%%%%%%%%%%%%%%%%%%%%%%%%%%%%%%%%%%%%%%%%%%%%%%%%%%%%%%%

\vskip0.3cm
{\large\bf Acknowledgments}
%\vskip0.3cm

The work of Yu.~K. is supported in part by the U.S. Department of
Energy under Grant No. DE-FG02-05ER41377. K.~T. would like to thank
RIKEN, BNL and the U.S. Department of Energy (Contract
No. DE-AC02-98CH10886) for providing the facilities essential for the
completion of this work.

%%%%%%%%%%%%%%%%%%%%%%%%%%%%%%%%%%%%%%%%%%%%%%%%%%%%%%%%%%%%%%%%%%%%%%%%%%%%

\end{document}